\documentclass[11pt]{iopart}
\usepackage{bm}
\usepackage{latexsym}
\usepackage{dcolumn}
\usepackage{graphicx}
\usepackage{epsfig}
\usepackage{psfrag}
\usepackage{iopams}

\def\be {\begin{equation}}
\def\ee {\end{equation}}
\def\bea {\begin{eqnarray}}
\def\eea {\end{eqnarray}}
\def\bse {\begin{subequations}}
\def\ese {\end{subequations}}
\def\bc {\begin{center}}
\def\ec {\end{center}}
\def\bfg {\begin{figure}}
\def\efg {\end{figure}}
\def\bi {\begin{itemize}}
\def\ei {\end{itemize}}
\def\nn {\nonumber}

\def\la {\label}
\def\le {\left}
\def\ri {\right}
\def\pa {\partial}
\def\fr {\frac}
\def\sq {\sqrt}

\def\b  {\beta}
\def\c  {\gamma}

\def\d  {\delta}

\def\f {\phi}

\def\k  {\kappa}

\def\L  {\Lambda}
\def\m  {\mu}
\def\n  {\nu}

\def\O  {\Omega}
\def\p  {\pi}
\def\r  {\rho}

\def\s {\sigma}

\def\vph {\varphi}
\def\vth {\vartheta}

\def\rx {\rho_{_X}}
\def\drx {\dot{\rho}_{_X}}
\def\rmat {\rho_{_m}}
\def\rmp {\rho_{_{0m}}}
\def\rcp {\rho_{_{0c}}}
\def\px {p_{_X}}
\def\wx {w_{_X}}

\def\df {\dot{\phi}}

\def\ta {\tilde{a}}
\def\tz {\tilde{z}}
\def\tilh {\tilde{h}}
\def\tH {\tilde{H}}

\begin{document}

\title{Crossing the cosmological constant barrier with kinetically interacting double quintessence}

\author{Sourav Sur}

\address{Dept. of Physics, University of Lethbridge \\
4401 University Drive, Lethbridge, Alberta, Canada T1K 3M4}

\eads{sourav.sur@uleth.ca}

\begin{abstract}
We examine the plausibility of crossing the cosmological constant ($\L$) barrier in a two-field quintessence
model of dark energy, involving a kinetic interaction between the individual fields. Such a kinetic interaction
may have its origin in the four dimensional effective two-field version of the Dirac-Born-Infeld action, that
describes the motion of a D3-brane in a higher dimensional space-time. We show that this interaction term could
indeed enable the dark energy equation of state parameter $\wx$ to cross the $\L$-barrier (i.e., $\wx = -1$),
keeping the Hamiltonian well behaved (bounded from below), as well as satisfying the condition of stability of
cosmological density perturbations, i.e., the positivity of the squares of the sound speeds corresponding to the
adiabatic and entropy modes. The model is found to fit well with the latest Supernova Union data and the WMAP
results. The best fit curve for $\wx$ crosses $-1$ at red-shift $z$ in the range $\sim 0.215 - 0.245$, whereas
the transition from deceleration to acceleration takes place in the range of $z \sim 0.56 - 0.6$. The scalar
potential reconstructed using the best fit model parameters is found to vary smoothly with time, while the dark
energy density nearly follows the matter density at early epochs, becomes dominant in recent past, and slowly
increases thereafter without giving rise to singularities in finite future.

\end{abstract}




\maketitle


\section{Introduction       \la{sec:intro}}

A variety of recent observational probes, including in particular the type Ia Supernovae (SN Ia)
\cite{perl,tonry,riess04,riess06,astier,davis,mik,kowal,rubin}, indicate that our universe has
entered in a phase of accelerated expansion in recent past, following an early decelerating
regime. Despite several alternative proposals, such as modified gravity \cite{modgrav} and the
averaging of cosmological inhomogeneities \cite{wilt}, the origin of this acceleration has
widely been attributed to a `mysterious' energy component, namely the dark energy (DE), which
constitutes about 72\% of the present universe. Moreover, the cosmic microwave background (CMB)
temperature fluctuation measurements by the Wilkinson Microwave Anisotropy Probe (WMAP)
\cite{wmap1,wmap3,wmap5} as well as the large scale red-shift data from the Sloan Digital Sky
Survey (SDSS) \cite{sdss} indicate that our universe is very nearly spatially flat, so that
spatial inhomogeneities may be neglected at large scales. Although the DE closely resembles a
positive cosmological constant $\L$, for which the DE equation of state (EoS) parameter $\wx
= -1$, there are some serious theoretical problems, such as {\it fine tuning} and {\it coincidence},
associated with $\L$ \cite{lambda}. Specifically, if the DE is supposed to be due to $\L$ and the
acceleration began only in recent past, then (i) what makes the DE density scale very small compared
to the Planck scale? and (ii) why is the DE density $\rx$ is of the order of the present critical
density $\rcp$ right now? Hence, there have been suggestions that the DE may be (more appropriately)
{\it dynamic} and can be modeled by one or more scalar field(s) originating from a fundamental theory.
Of major interest are the DE models developed in the framework of quintessence and tracker fields
\cite{quin,tracker}, k-essence \cite{k-ess1,k-ess2}, Chaplygin gas \cite{chap} etc., (see
\cite{de-rev} for extensive reviews). However, in many of these models the value of $\wx$ is always
restricted to be $\geq -1$, which is not desirable for a consistent statistical fit with the
observational data. In fact, even with the presumption that $\wx$ is a constant, the recent WMAP
five year data, combined with with those for SN Ia and baryon acoustic oscillation (BAO) peaks,
constrain the value of $1+\wx$, to be between $-0.14$ and $0.12$, at 95\% confidence level (CL)
\cite{wmap5}. For a time-varying DE, the same data constrain the value $w_{_{0X}}$ of the DE EoS
parameter at the present epoch (i.e., at red-shift $z = 0$) to be between $-1.33$ and $0.79$ (at
95\% CL) \cite{wmap5}. Since in the distant past, the value of a variable $\wx$ must have to be
$\gg -1$ (so that the universe had a decelerated expansion and structures were formed), there is
a fair plausibility that one (or possibly more) transition(s) from $\wx > -1$ to $\wx < -1$ (or
vice versa) could have been taken place in the recent course of evolution of the DE, and at present
$\wx = w_{_{0X}} < -1$.

The crossing of the cosmological constant barrier ($\wx = -1$) can, most simply, be achieved
in the so-called {\it quintom} scenario \cite{quintom}, where there are two (or more) scalar
fields, (at least) one of which is of `phantom' nature, i.e., carries a wrong sign in front of
the kinetic term in the Lagrangian \cite{phantom}. Such a phantom field is quantum mechanically
unstable \cite{cline} and also gives rise to singularities in finite future \cite{cald,noj}.
Moreover, classical instabilities could arise as the dominant energy condition gets violated
in the models involving the phantom fields \cite{cht}. Attempts have therefore been made to
circumvent the problem of $\wx = -1$ crossing in various alternative ways. Notable among
these are the scalar-tensor models \cite{polarski}, brane-world models \cite{sahni}, multi-field
k-essence models \cite{chim,sssd}, modified gravity models \cite{modgravPDL}, string-inspired
dilatonic ghost condensate models \cite{dil}, quantum-corrected Klein-Gordon models with quartic
potential \cite{onemli}, coupled DE models \cite{coupDE}, H-essence (complex scalar) models
\cite{wei}, etc. However, apart from a very few exceptions (such as the scalar-tensor models
\cite{polarski}, or models where the kinetic term abruptly flips sign due to some extraordinary
nature of the potential \cite{andrianov}) the $\wx = -1$ crossing is hard to be realized with a
single-field DE. Even in the case of a single-field k-essence DE, with a generic non-linear
dependence of the Lagrangian on the kinetic term, such a crossing either leads to instabilities
against cosmological perturbations or is realized by a discrete set of phase space
trajectories\footnote{Of course, there are exceptions as well, see for example ref.
\cite{odint}.} \cite{vik}. In multi-field DE models, however, the $\wx = -1$ crossing could
be made possible, as is shown for example in refs. \cite{chim,sssd}, although the field
configuration may be severely constrained by the criterion of stability, i.e., the square of
the effective speed of propagation of cosmological perturbations should be positive definite
\cite{sssd}.

In this paper we explore the plausibility of the $\L$-barrier crossing in the framework of a
two-field quintessence model with a kinetic interaction between the individual fields. Such
a model may be looked upon as a specialization of a more general (interacting) multi-field
k-essence scenario, which involves non-canonical (higher order) kinetic terms for the scalar
fields \cite{k-ess1,sssd,lang1,lang2,lang}. Moreover, the kinetically interacting double
quintessence (KIDQ) Lagrangian may, under certain approximations, be derived from the four
dimensional effective two-field version of the Dirac-Born-Infeld (DBI) Lagrangian describing
the evolution of D3-branes in higher dimensional string theoretic manifolds \cite{dbi}. The
biggest advantage with such a Lagrangian, compared to those in other $\L$-barrier crossing
multiple k-essence models \cite{chim,sssd}, is that the total DE Hamiltonian consists of a
positive definite kinetic part, which ensures that it is bounded from below and the model is
quantum mechanically consistent. Stability against cosmological density perturbations further
requires the squares of the effective (sound) speeds of propagation of the adiabatic and
entropy modes to be positive definite as well. For the DBI multiple scalar fields in
homogeneous cosmological backgrounds, both these sound speeds turn out to be the same,
implying isotropic propagation of the adiabatic and entropy modes \cite{lang1,lang2}.
Assuming this result to hold approximately for KIDQ (which is an approximation to the DBI
two-scalar scenario), we find the square of the effective (isotropic) sound speed to be
positive definite, ensuring the stability of the KIDQ model\footnote{More precisely,
however, there is a splitting between the propagation speeds of the adiabatic and the
entropy modes, when the KIDQ is taken to be an exact theory (not an approximation to DBI).
This we find in a subsequent paper \cite{ss-prep1} (in preparation) by carrying out the
stability analysis for KIDQ, following the general formalism worked out in refs.
\cite{lang2,lang} in the context of multi-field DBI and k-inflation. The squares of the
propagation speeds turn out to be positive definite anyway.}.

We consider certain specific ansatze to solve for the KIDQ field equations, and obtain the
condition under which the $\wx = -1$ line could be crossed in some regime. In choosing the
ansatze, we particularly emphasize on the following:
\begin{description}
\item (i) the kinetic energy densities of the interacting scalar fields should always be
positive definite,
\item (ii) the DE density should be less but not very smaller than the matter density at
early epochs, and should dominate the latter at late times, and
\item (iii) the DE density should not grow rapidly with increasing scale factor $a$ (i.e.,
decreasing red-shift $z$) and reach to abnormally high values in finite future.
\end{description}
These are important in order to avoid (i) ghosts or phantoms, (ii) coincidence or
fine-tuning related problems, and (iii) occurance of future singularities, respectively.

We then constrain the parameters of the model with the latest Supernova Ia data compiled
in ref. \cite{kowal}, viz., the 307 Union data-set, as well as with the WMAP 5-year
\cite{wmap5} update of the CMB-shift parameter $\mathcal{R}$ and the scalar spectral
index $n_s$, which determines the BAO peak distance parameter $\mathcal{A}$ from the
SDSS luminous red galactic distribution \cite{sdss}. After uniformly marginalizing over
the Hubble constant $H_0$, we obtain good fits of the model with the data. The minimized
value of the total $\chi^2$ (SN+CMB+BAO) is found to be $\simeq 311$, which is
better than the minimized $\chi^2 (\simeq 313)$ found with the Union data-set in ref.
\cite{rubin} for the cosmological constant DE coupled with cold dark matter -- the
so-called $\L$CDM model. The best fit values of the parameters of our KIDQ model indicate
that the crossing from $\wx > -1$ to $\wx < -1$ takes place at a red-shift range
$0.215 \leq z_c \leq 0.245$, whereas the transition from the decelerated regime to the
accelerated regime occurs in the range $0.562 \leq z_t \leq 0.603$. At the present
epoch ($z = 0$), the best fit values of the matter density parameter and the DE EoS
parameter, are respectively found to lie within the ranges $0.279 \leq \O_{0m} \leq
0.281$ and $-1.123 \leq w_{_{0X}} \leq -1.077$. All these results are fairly in agreement
with those found with other model-independent or model-dependent parameterizations of the
DE in the literature \cite{cpl,alam,visser,param,pdlc}.

Finally, we integrate the scalar field equations of motion and reconstruct the
interacting double quintessence potential using the best fit model parameters. We show
that the reconstructed potential has a smooth dependence (i.e., without any discontinuity
or multi-valuedness) on the scale factor $a$. We work out the approximate analytic expressions
for the potential as function of the scalar fields, and find that they also exhibit the same
smooth nature at early and late stages of the evolution of the universe.

This paper is organized as follows: in sec. \ref{sec:gf} we describe the general framework
of the multi-scalar (k-essence) DE scenario, following the formalism shown in refs.
\cite{lang2,lang} in the context of multi-field k-inflation. In sec. \ref{sec:model} we
emphasize on a special case which involves two quintessence type of scalar fields with
canonical kinetic terms in the Lagrangian, and with a specific kinetic interaction between
the individual fields. Assuming suitable ansatze for the solutions of the field equations
we work out the condition under which the cosmological constant barrier $\wx = -1$ could
be crossed, and find the expression for the Hubble parameter maintaining this condition.
In sec. \ref{sec:modelfit} we fit our KIDQ model with the $307$ Union SN Ia data \cite{kowal},
combined with the CMB+BAO results from WMAP and SDSS, to obtain the DE density and EoS
profiles. In sec. \ref{sec:potential} we use the best fit values of the model parameters to
reconstruct phenomenologically the interacting double quintessence potential and determine
the temporal variations of the scalar fields. We also work out the approximate analytic
functional forms of the potential in terms of the scalar fields, at early and late stages of
the evolution of the universe. In sec. \ref{sec:concl}, we conclude with a summary and some
open questions. In the Appendix, we show how the KIDQ action, that we consider, could be
derived from the two-field DBI action under certain approximations.

\section{General Formalism        \la{sec:gf}}

Let us consider the following action, in $(3+1)$ dimensions, for gravity minimally coupled
with matter fields and $N$ number of kinetically interacting (k-essence) scalar fields $\f^I$
($I = 1, \dots, N$):
\be \la{gen-action}
S = \int d^4 x \sq{-g} \le[\fr R {2 \k^2} ~+~ {\mathcal L}_m ~+~ P \le(X^{IJ}, \f^K\ri) \ri] \, ,
\ee
where $\k^2 = 8 \p G$ is the gravitational coupling constant, ${\mathcal L}_m$ is the Lagrangian
density for matter, that is considered to be pressureless dust. $P (X^{IJ}, \f^K)$ is the
multi-scalar Lagrangian density, with
\be \la{k-term}
X^{IJ} ~=~ -~ \fr 1 2~ g^{\m\n}~ \pa_\m \f^I~ \pa_\n \f^J \, , \qquad (I,J = 1, \dots, N) \, ,
\ee
describing the kinetics of the scalar fields \cite{lang2}.

In a spatially flat Friedmann-Robertson-Walker (FRW) background, with line element
\be \la{RW}
ds^2 = - dt^2 + a^2 (t) \le[dr^2 + r^2 \le(d\vth^2 + \sin^2 \vth d\vph^2\ri)\ri]
\ee
the above expression for $X^{IJ}$ reduces to
\be \la{k-term1}
X^{IJ} ~=~ X^{JI} ~=~ \fr 1 2~ \df^I ~ \df^J ~=~ \fr{a^2 H^2} 2 ~\f^{'I} ~\f^{'J} \, ,
\ee
where the dot denotes time derivative ($d/dt$) and the prime denotes derivative ($d/da$) with
respect to the scale factor $a$, which has been normalized to unity at the present epoch $t =
t_0$. $H \equiv \dot{a}/a$ is the Hubble parameter.

The Friedmann equations and the scalar field equations of motion are given by
\be \la{FRW-eom}
\fl \qquad H^2 \equiv \le(\fr{\dot{a}}a\ri)^2 = \fr{\k^2} 3 \le(\rmat + \rx\ri) \quad , \qquad
\dot{H} \equiv \fr{\ddot{a}}a - \fr{\dot{a}^2}{a^2} = - \fr{\k^2}2 \le[\rmat +
\le(\rx + \px\ri)\ri] \, ,
\ee
\be \la{multi-sf-eom}
\fr d {dt} \le(a^3 ~\fr {\pa P}{\pa X^{IJ}}~ \df^J\ri) =~ a^3~ \fr {\pa P}{\pa \f^I} \, ,
\ee
where $\rmat$ is the energy density of matter in the form pressureless dust, and $\rx, \px$
are the multi-field dark energy density and pressure, given respectively as
\be \la{multi-ed-pr}
\rx ~=~ 2 X^{IJ} \fr {\pa P}{\pa X^{IJ}} ~-~ P \quad, \qquad \px ~=~ P \, .
\ee

Assuming that there is no mutual interaction between matter and dark energy, the Friedmann
equations (\ref{FRW-eom}) integrate to give $\rmat = \rmp a^{-3}$, where $\rmp$ is the matter
density at the present epoch ($t = t_0, a = 1$). One also has the continuity equation for
the dark energy
\be \la{cont-eq}
\drx = - 3 H \le(\rx + \px\ri) \qquad \Rightarrow \qquad \rx' = - \fr 3 a \le(\rx + \px\ri) \, .
\ee

From the Friedmann equations (\ref{FRW-eom}) one obtains the expressions for the DE EoS parameter
$\wx$, the total EoS parameter $w$, and the deceleration parameter $q$:
\bea
\la{de-eos}
\wx &=& \fr{\px}{\rx} ~=~ - 1 ~+~ \fr {2 X^{IJ}} {\rx} ~\fr {\pa P}{\pa X^{IJ}} \, ,  \\
\la{tot-eos}
w &=& \fr{\px}{\rmat + \rx} ~=~ \wx \le(1 - \fr{\O_{0m}}{{\tilde H}^2 a^3}\ri) \, , \\
\la{decel}
q &\equiv& - \fr{\ddot{a}}{a H^2} ~=~ \fr{1 + 3 w} 2 \, ,
\eea
where $\rcp = 3 H_0^2/\k^2$ is the present critical density; $H_0$ being the value of $H$ at
the present epoch ($t = t_0$).
\be \la{norm-hub}
\tH ~\equiv~ \fr H {H_0} ~=~ \sq{\fr{\rx}{\rcp} ~+~ \fr{\O_{0m}}{a^3}} \, ,
\ee
is the normalized Hubble parameter and $\O_{0m} = \rmp/\rcp$ is the present matter density
parameter.

The transition from the decelerating regime to the accelerating regime takes place when
the deceleration parameter $q$ changes sign, i.e., the total EoS parameter $w$ becomes less
than $- 1/3$, by Eq. (\ref{decel}), and the DE EoS parameter $\wx$ is further less, by Eq.
(\ref{tot-eos}). The crossing from $\wx > -1$ to $\wx < -1$, on the other hand, requires
a flip of sign of the quantity $X^{IJ} \pa P/\pa X^{IJ}$, presuming that the DE density
$\rx$ is positive definite. In the next section, we examine the plausibility of such a
crossing by considering for simplicity a model involving only two fields ($N = 2$) with
usual (canonical) kinetic terms (quintessence type), but with a specific type of kinetic
interaction, which could have its origin in the two-field DBI action, as we show in the
Appendix.

\section{Kinetically interacting double quintessence \la{sec:model}}

Let us take into account the following special form of the Lagrangian density for the
DE, consisting of only two scalar fields:
\be \la{de-action}
P ~=~ \d_{IJ} X^{IJ} ~-~ \c \sqrt{1 ~-~ \fr{\b} 2 \le(\d_{I,J-1} ~+~ \d_{I-1,J}\ri)
X^{IJ}} ~-~ V (\f^I) \, ,
\ee
where $\b, \c$ are positive constants, $V (\f^I)$ is the scalar potential, and the indices
$I, J$ run for $1, 2$. Denoting the  two fields as $\f^I \equiv (\f, \xi)$, we can re-write
the above Lagrangian as
\be \la{de-action1}
\fl \qquad P ~=~ \fr{\df^2} 2 ~+~ \fr{\dot{\xi}^2} 2 ~-~ \c ~Q (\df, \dot{\xi}) ~-~ V (\f,
\xi) \, , \qquad \mbox{where} \quad Q (\df, \dot{\xi}) ~=~ \sqrt{1 ~-~ \fr{\b} 2 ~ \df
\dot{\xi}} \, .
\ee
This implies that the scalar fields $\f$ and $\xi$ have usual (canonical) kinetic energy
densities (given respectively by the first two terms on the right hand side), and therefore
are similar to ordinary quintessence fields. However they have a mutual kinetic interaction
of a specific form proportional to $Q (\df, \dot{\xi})$, given above, which may originate
from the two-scalar DBI action, approximated for $\b \ll 1$ and $\c \gg 1$ (but $\c^{-1}
\ll \b$) as shown in the Appendix.

The dark energy pressure $\px$ is equal to $P$ in Eq. (\ref{de-action1}), whereas the
expression (\ref{multi-ed-pr}) for the dark energy density reduces to
\be \la{de-den}
\rx ~=~ \fr{\df^2} 2 ~+~ \fr{\dot{\xi}^2} 2 ~+~ \fr{\c} {Q (\df, \dot{\xi})} ~+~ V (\f, \xi) \, .
\ee
The DE equation of state parameter $\wx$, Eq. (\ref{de-eos}), now takes the form
\be \la{de-eos1}
\wx ~=~ \fr{\px}{\rx} ~=~ - 1 ~+~ \fr 1 {\rx} \le[\df^2 ~+~ \dot{\xi}^2 ~+~ \fr {\b~ \c ~ \df~
\dot{\xi}}{2 Q (\df, \dot{\xi})}\ri] \, .
\ee

The presumption that the parameter $\b \ll 1$, is in support of the positivity of the term under
the square root in the expression for $Q (\df, \dot{\xi})$ given in Eq. (\ref{de-action1}). That
is, the requirement $Q^2 (\df, \dot{\xi}) > 0$ for the validity of the model, could be fulfilled
when $\b \ll 1$, even if $\df$ and $\dot{\xi}$ vary fairly rapidly with time and the product $\df
\dot{\xi} > 0$. Considering further, $Q (\df, \dot{\xi})$ itself to be positive, the kinematical
part of the DE density, given by the first three terms (kinetic energy densities of the fields
plus their kinetic interaction) on the right hand side of Eq. (\ref{de-den}), remains positive
definite. As such, the total DE Hamiltonian is bounded from below and the model is quantum
mechanically consistent. Moreover, since $\b, \c$ and $Q (\df, \dot{\xi})$ are all positive, it
follows from Eq. (\ref{de-eos1}) that, $\wx < -1$ (in some regime) necessarily implies the product
$\df \dot{\xi} < 0$. In other words, the condition for the crossing of the $\wx = -1$ barrier at a
particular epoch, is that one of the two fields ($\f, \xi$) must fall off with time, whereas the
other one should increase with time.

Now, using Eqs. (\ref{de-den}), (\ref{de-eos1}) and the continuity equation (\ref{cont-eq}),
one obtains the following expression for the potential $V$ as a function of the scale factor
$a$:
\be \la{pot}
\fl \quad V (a) ~=~ - \fr{\df^2 (a) + \dot{\xi}^2 (a)} 2 ~-~ \fr{\c} {Q (a)} ~+~ \L ~-~ 3 \int^a
\fr{d\ta}{\ta} \le[\df^2 (\ta) + \dot{\xi}^2 (\ta) + \fr{\b \c ~\df (\ta) ~\dot{\xi} (\ta)}
{2~ Q (\ta)}\ri] ,
\ee
where $\L$ is an integration constant. Plugging Eq. (\ref{pot}) back in Eq. (\ref{de-den})
we get the DE density $\rx$ as a function of $a$:
\be \la{de-den1}
\rx (a) ~=~ \L ~-~ 3 \int^a \fr{d\ta}{\ta} \le[\df^2 (\ta) + \dot{\xi}^2 (\ta) +
\fr{\b \c ~\df (\ta) ~ \dot{\xi} (\ta)} {2~ Q (\ta)}\ri]  \, .
\ee
One may note that Eq. (\ref{pot}) could also have been obtained by using the scalar field
equations of motion (\ref{multi-sf-eom}), which in the present scenario reduce to
%
\bea \la{sf-eom}
\fr d {dt} \le[a^3 \le(\df ~+~ \fr{\b \c ~\dot{\xi}}{4 Q (\df, \dot{\xi})}\ri)\ri]
=~ a^3 ~\fr{\pa V}{\pa \f} \, , \nn \\
\fr d {dt} \le[a^3 \le(\dot{\xi} ~+~ \fr{\b \c ~\df}{4 Q (\df, \dot{\xi})}\ri)\ri] =~ a^3
~\fr{\pa V}{\pa \xi} \, .
\eea
%

Under a dimensional re-scaling:
\be \la{rescale}
\f \leftrightarrow \fr{\f}{\sq{\rcp}} \, , \quad \xi \leftrightarrow \fr{\xi}{\sq{\rcp}} \, ,
\quad \L \leftrightarrow \fr{\L}{\rcp} \, , \quad \b \leftrightarrow \b \rcp \, , \quad
\c \leftrightarrow \fr{\c}{\rcp} \, ,
\ee
the DE density, pressure, and the scalar potential change as
\be \la{de-rescale}
\rx \leftrightarrow \fr{\rx}{\rcp} \, , \qquad \px \leftrightarrow \fr{\px}{\rcp} \, , \qquad
V \leftrightarrow \fr{V}{\rcp} \, ,
\ee
while all the above equations (\ref{de-den}) - (\ref{sf-eom}) remain invariant. On the other hand,
the expression (\ref{norm-hub}) for the normalized Hubble parameter reduces to
\be \la{norm-hub1}
\fl \quad \tH^2 (a) ~=~ \rx (a) ~+~ \fr{\O_{0m}}{a^3} ~=~ \L ~+~ \fr{\O_{0m}}{a^3} ~-~
3 \int^a \fr{d\ta}{\ta} \le[\df^2 (\ta) + \dot{\xi}^2 (\ta) + \fr{\b \c ~\df (\ta) ~ \dot{\xi}
(\ta)} {2~ Q (\ta)}\ri]  .
\ee

Let us now consider the following ansatze for the kinetic energy densities of the scalar fields:
%
\bea \la {ansz}
\r_{_{\f}}^{K} (a) &=& \fr 1 2~ \df^2 (a) ~=~ \fr 1 2 \le[f (a) ~+~ \sq{f^2 (a) ~-~
k^2}\ri] \, ,  \nn \\
\r_{_{\xi}}^{K} (a) &=& \fr 1 2~ \dot{\xi}^2 (a) ~=~ \fr 1 2 \le[f (a) ~-~ \sq{f^2 (a)
~-~ k^2}\ri] \, ,
\eea
%
where $f (a)$ is taken to be a positive definite and well-behaved function of $a$, $k$ is a
positive constant, and $f (a) > k$ at all epochs. Eqs. (\ref{ansz}) imply that
\be \la{ansz1}
\df^2 ~+~ \dot{\xi}^2 ~=~ 2 f (a) \, , \quad \mbox{and} \qquad \df~ \dot{\xi} ~=~ \pm k \, .
\ee
We choose to take $\df \dot{\xi} = - k$, so that the DE EoS parameter $\wx$, Eq. (\ref{de-eos1}),
could be made less than $-1$ in some regime. Moreover, this choice guarantees the positivity of
the square of the kinetic interaction, which now reduces to a constant:
\be \la{Q-sq}
Q^2 ~=~ 1 ~+~ \fr{\b k} 2 \, .
\ee
The expressions for the time derivatives of the scalar fields are given by
\be \la{sf-deriv}
\fl \quad \df (a) ~=~ \fr{\sq{f(a) - k} + \sq{f(a) + k}}{\sq{2}} \, , \quad
\dot{\xi} (a) ~=~ \fr{\sq{f(a) - k} - \sq{f(a) + k}}{\sq{2}} \, ,
\ee
whereas from Eqs. (\ref{de-eos1}), (\ref{de-den1}) and (\ref{pot}), we respectively obtain the
following expressions for the DE EoS parameter and density, and the scalar potential:
\bea
\la{de-eos2}
\wx (a) &=& - 1 ~+~ \fr 1 {\rx (a)} \le[2 f (a) ~-~ \fr{\b \c k}{2 Q}\ri] \, , \\
\la{de-den2}
\rx (a) &=& \L ~+~ \fr{3 \b \c k}{2 Q} ~\ln a ~-~ 6 \int^a \fr{f (\ta)}{\ta} d\ta \, , \\
\la{pot1}
V (a) &=& \L ~+~ \fr{3 \b \c k}{2 Q} ~\ln a ~-~ f (a) ~-~ \fr{\c} Q ~-~ 6 \int^a
\fr{f (\ta)}{\ta} d\ta \, .
\eea

Let us now assume a specific form of the function $f(a)$, given by
\be \la{f-ansz}
f (a) ~=~ A a^{- \n} ~+~ k \, , \quad \mbox{where} \qquad A > 0 \, , \quad 0 < \n < 3 \, ,
\ee
so that the criterion $f (a) > k > 0$ is automatically satisfied. Furthermore, $0 < \n < 3$
ensures that $f (a)$, and hence the kinetic energy densities $\fr 1 2 \df^2$ and $\fr 1 2
\dot{\xi}^2$ of the scalar fields, fall off with increasing values of the scale factor $a$.
However, these fall offs are not faster than that of the matter density ($\rmat \sim 1/a^3$).
This is essential in order that the quantities $\fr 1 2 \df^2$ and $\fr 1 2 \dot{\xi}^2$,
which compose the total DE density $\rx$, come to dominate $\rmat$ at late times, i.e., for
large values of $a$.

Eqs. (\ref{sf-deriv}) reduce to
\be \la{sf-deriv1}
\df (a) ~=~ \fr{\sq{A a^{-\n}} + \sq{A a^{- \n} + k}}{\sq{2}} \, , \quad
\dot{\xi} (a) ~=~ \fr{\sq{A a^{-\n}} - \sq{A a^{- \n} + k}}{\sq{2}} \, ,
\ee
and the Eqs. (\ref{de-eos2}) - (\ref{pot1}), for $\wx, \rx$ and $V$, take the form
\bea
\la{de-eos3}
\wx (a) &=& - 1 ~+~ \fr 2 {\rx (a)} \le(A a^{-\n} ~-~ B\ri) \, , \\
\la{de-den3}
\rx (a) &=& \fr{6 A}{\n} a^{-\n} ~+~ 6 B \ln a ~+~ \L \, , \\
\la{pot2}
V (a) &=& V_0 ~+~ \le(\fr 6 {\n} - 1\ri) A \le(a^{-\n} - 1\ri) +~ 6 B \ln a \, ,
\eea
where we have defined
\be \la{B-def}
B ~=~ k \le(\fr{\b \c}{4 Q} ~-~ 1\ri) =~ \mbox{constant} \, ,
\ee
and $V_0$ is the value of the scalar potential $V$ at the present epoch ($t = t_0,
a = 1$):
\be \la{V0}
V_0 ~=~ \le(\fr 6 {\n} - 1\ri) A ~+ \le(\L ~-~ k ~-~ \fr{\c} Q\ri) \, .
\ee
From Eqs. (\ref{norm-hub1}) and (\ref{de-den3})  one also obtains the following
expression for the normalized Hubble rate $\tH = H/H_0$:
\be \la{norm-hub2}
\tH^2 (a) ~=~ \fr{6 A}{\n a^\n} ~+~ \fr{\O_{0m}}{a^3} ~+~ 6 B \ln a ~+~ \L \, .
\ee
At $a = 1$ (present epoch), $\tH = 1$, whence
\be \la{CC}
\L ~=~ 1 ~-~ \O_{0m} ~-~ \fr{6 A}{\n} \, ,
\ee
and the above expression (\ref{norm-hub2}) reduces to
\be \la{norm-hub3}
\tH^2 (a) ~=~ 1 ~+~ \fr{6 A}{\n a^\n} \le(1 - a^\n\ri) +~ \fr{\O_{0m}}{a^3}
\le(1 - a^3\ri) +~ 6 B \ln a \, .
\ee

In the next section, we fit this Eq. (\ref{norm-hub3}) with the latest Supernova
Ia data \cite{kowal}, as well as with the CMB+BAO results from WMAP and SDSS
\cite{wmap5,sdss}, and determine the DE density and EoS profiles over the red-shift
range that is probed.

\section{Observational constraints  \la{sec:modelfit}}

We perform a $\chi^2$ analysis so as to constrain the model parameters $A, B$ and $\O_{0m}$, for
two specific choices of the index $\n$ ($= 1, 2$) in the ansatze (\ref{f-ansz}). The SN Ia Union
data-set \cite{kowal}, which we use, consists of $307$ most reliable data points that range up
to red-shift $z = (1/a - 1) \sim  1.7$, and include large samples of SN Ia from older data-sets
\cite{perl,tonry,riess04,riess06}, high-$z$ Hubble Space Telescope (HST) observations and the SN
Legacy Survey (SNLS) \cite{astier}.

The SN Ia data provide the observed distance modulus $\m_{obs} (z_i)$, with the respective
$1\s$ uncertainty $\s_i (z_i)$, for SN Ia located at various red-shifts $z_i , (i = 1, \dots,
307)$. The $\chi^2$ for the SN observations is, on the other hand, expressed as
\be \la{chisq-SN}
\chi_{_{SN}}^2 (\m_0; \O_{0m}, A, B) ~=~ \sum_{i=1}^{307} \fr{\le[\m_{obs} (z_i) -
\m (z_i)\ri]^2}{\s_i^2 (z_i)} \, ,
\ee
where
\be \la{dist-mod}
\m (z_i) ~=~ 5 \log_{10} \le[D_L (z_i)\ri] + \m_0 \, ,
\ee
is the theoretical distance modulus.
\be \la{lum-dist}
D_L (z_i) ~=~ \le(1 + z_i\ri) \int_0^{z_i} \fr{d \tz_i}{{\tilde H} (\tz_i; \O_{0m}, A, B)} \, ,
\ee
is the Hubble free luminosity distance in terms of the parameters $\le(\O_{0m}, A, B\ri)$,
and
\be \la{nuis}
\m_0 ~=~ 5 \log_{10} \le[\fr{H_0^{-1}}{Mpc}\ri] +~ 25 ~=~ 42.38 ~-~ 5 \log_{10} h \, ,
\ee
$h$ being the Hubble constant $H_0$ in units of $100$ Km s$^{-1}$ Mpc$^{-1}$. The parameter $\m_0$
is a nuisance parameter, independent of the data points, and has to be uniformly marginalized over
(i.e., integrated out). For such a marginalization one may follow the procedure shown in refs.
\cite{sssd,peri,wei08}, where $\chi_{_{SN}}^2$ is first expanded suitably in terms of $\m_0$. Then
one finds the value of $\m_0$ for which such an expanded form of $\chi_{_{SN}}^2$ is minimum.
Substituting this value of $\m_0$ back in $\chi_{_{SN}}^2$, finally enables one to perform the
minimization of the resulting expression with respect to the parameters $\le(\O_{0m}, A, B\ri)$,
in order to determine the values of the latter best fit with the SN Ia observations.

The CMB shift parameter $\mathcal{R}$, that relates the angular diameter distance to the last
scattering surface (at red-shift $z_{ls}$) with the co-moving sound horizon scale at recombination
and the angular scale of the first acoustic peak in the CMB temperature fluctuations power spectrum
\cite{sdss,cmb-R}, is given by
\be \la{R}
\mathcal{R} (z_{\star}) ~=~ \O_{0m}^{1/2} ~ \int_0^{z_{\star}} \fr{d\tz}{{\tilde H} (\tz; \O_{0m},
A, B)} \, ,
\ee
where $z_{\star}$ is the red-shift of recombination. The WMAP five year data \cite{wmap5} updates
$z_{\star} = 1090.04 \pm 0.93$ and the observed shift parameter $\mathcal{R}_{obs} (z_{\star}) =
1.710 \pm 0.019$. The $\chi^2$ for the CMB observations is given by
\be \la{chisq-CMB}
\chi_{_{CMB}}^2 ~=~ \fr{\le(\mathcal{R}_{obs} ~-~ \mathcal{R}\ri)^2}{\s_R^2} \, ,
\ee
where $\s_R$ is the $1\s$ error in the WMAP data \cite{wmap5}.

Now, the scalar spectral index $n_s$, which determines the observed value of the BAO peak distance
parameter $\mathcal{A}_{obs}$ from the distribution of the SDSS luminous red galaxies \cite{sdss}
through the relation $\mathcal{A}_{obs} = 0.469 \le(n_s/0.98\ri)^{-0.35} \pm 0.017$, is updated by
the WMAP five year data as $n_s = 0.960 \pm 0.013$ (see the first ref. of \cite{wmap5}, see also
ref. \cite{wei08}). The theoretical expression for the distance parameter is, on the other hand,
given by
\be \la{A}
\mathcal{A} ~=~ \O_{0m}^{1/2} \le[\fr 1 {z_b \sq{\tH (z_b)}} ~ \int_0^{z_b} \fr{d\tz}
{{\tilde H} (\tz; \O_{0m}, A, B)}\ri]^{2/3} \, ,
\ee
where $z_b = 0.35$. The $\chi^2$ for the BAO observations is expressed as
\be \la{chisq-BAO}
\chi_{_{BAO}}^2 ~=~ \fr{\le(\mathcal{A}_{obs} ~-~ \mathcal{A}\ri)^2}{\s_A^2} \, ,
\ee
$\s_A$ being the $1\s$ error in the SDSS data \cite{sdss}.

The total $\chi^2$, which needs to be minimized in order to determine the likelihood of the
model parameters $\le(\O_{0m}, A, B\ri)$ with the entire SN+CMB+BAO data, is thus given as
\be \la{chisq-tot}
\chi_{total}^2 ~=~ \chi_{_{SN}}^2 ~+~ \chi_{_{CMB}}^2 ~+~ \chi_{_{BAO}}^2 \, .
\ee
Of course, $\chi_{_{SN}}^2$ has already been minimized with respect to the nuisance parameter
$\m_0$, Eq. (\ref{nuis}), by the process discussed above.

\begin{table*}[!htb]
\la{table:1}
\begin{center}
\begin{tabular}{|c||c|c|c||c|}
\hline
Index &\multicolumn{3}{c|}{Best fit model parameters}&\multicolumn{1}{c|}{Minimized} \\
\cline{2-4}
$\n$&$~\O_{0m}~$&$~A~$&$~B~$&$~\chi_{total}^2~$ \\
\hline\hline
& & & &  \\
$1$&$~0.2790~$&$~0.2062~$&$~0.2506~$&$~311.07~$ \\
& & & &  \\
$2$&$~0.2816~$&$~0.0505~$&$~0.0782~$&$~311.27~$ \\
\hline
\end{tabular}
\end{center}
\caption{\small Values of the parameters $\le(\O_{0m}, A, B\ri)$ of the model, best fit with
SN+CMB+BAO observations, and the minimized total $\chi^2$, for the choices $\n = 1, 2$.}
\end{table*}

\begin{figure}[!htb]
\begin{center}
\includegraphics[width=12.5cm,height=12.5cm]{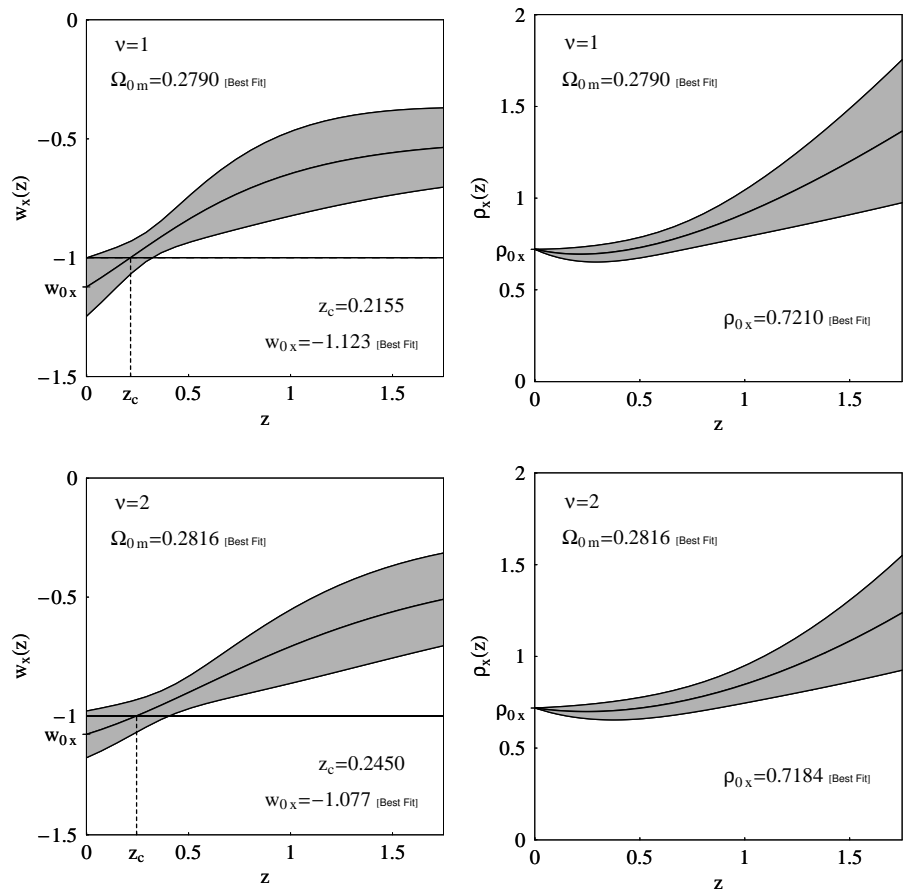}
\end{center}
\caption{\small Evolution of $\wx (z)$ and $\rx (z)$ (best fit with the SN+CMB+BAO observations)
throughout the red-shift range $0 \leq z \leq 1.75$) alongwith the corresponding $1\s$ error
(shaded) regions, are shown for the choices $\n = 1$ (upper panels) and $\n = 2$ (lower panels).
The point $z_c$ denotes the red-shift at which the $\wx = -1$ line is crossed, and $w_{_{0X}},
\r_{_{0X}}$ are respectively the values of $\wx, \rx$ at the present epoch ($z = 0$).}
\la{fig:1}
\end{figure}

For two specific choices of the index $\n$ ($= 1, 2$), that appears in the ansatz (\ref{f-ansz}),
the best fit values the parameters $\le(\O_{0m}, A, B\ri)$, as well as the minimized value of
$\chi_{total}^2$, are shown in the table 1. Fig. \ref{fig:1} shows the evolution of $\wx (z)$
and $\rx (z)$ (alongwith the corresponding $1\s$ errors) throughout the entire red-shift range
$0 \leq z \leq 1.75$ of the available data, for both the choices of $\n$. The maximum likelihood
of the present value $w_{_{0X}}$ of the DE EoS parameter is found to be $- 1.123$ for $\n = 1$
and $- 1.077$ for $\n = 2$. Both these values are well within the limits, viz., $-1.33 \leq
w_{_{0X}} \leq 0.79$, obtained in model-independent estimates with the SN+CMB+BAO data in ref.
\cite{wmap5}. On the other hand, the red-shift $z = z_c$ at which the best fit $\wx$ makes a
transition from a value $> -1$ to a value $< -1$ is found to be $0.2155$ for $\n = 1$ and $0.2450$
for $\n = 2$. However, $\wx$ stays well below zero even for $z = 1.75$, implying that DE is varying
slowly with red-shift. The above values of $z_c$ also agree fairly well with other independent
studies \cite{pdlc}. The best fit DE density at the present epoch, $\r_{_{0X}}$, is found to be
equal to $0.7210$ for $\n = 1$ and $0.7184$ for $\n = 2$. Remembering the dimensional re-scaling
of the DE density, viz., $\rx \leftrightarrow \rx/\rcp$, that we have performed earlier in Eq.
(\ref{de-rescale}), one may note that the $\r_{_{0X}}$ shown in Fig. \ref{fig:1} is identical
with the present DE density parameter $\O_{0X} = \r_{_{0X}}/\rcp$ (by virtue of the dimensional
re-scaling). In other words, since the DE density $\rx$ is effectively measured in units of
the present critical density $\rcp$, one has $\r_{_{0X}} \equiv \O_{0X}$. It may also be noted
that the sum of the best fit $\O_{0m}$ and the best fit $\O_{0X}$ is exactly equal to $1$ (for
both $\n = 1$ and $\n = 2$), as it should be in accord with our prior assumption of the spatial
flatness of the metric. This therefore proves the correctness of the $\chi^2$-fitting of the
model with the observational data.

\begin{figure}[!htb]
\begin{center}
\includegraphics[width=15cm,height=10cm]{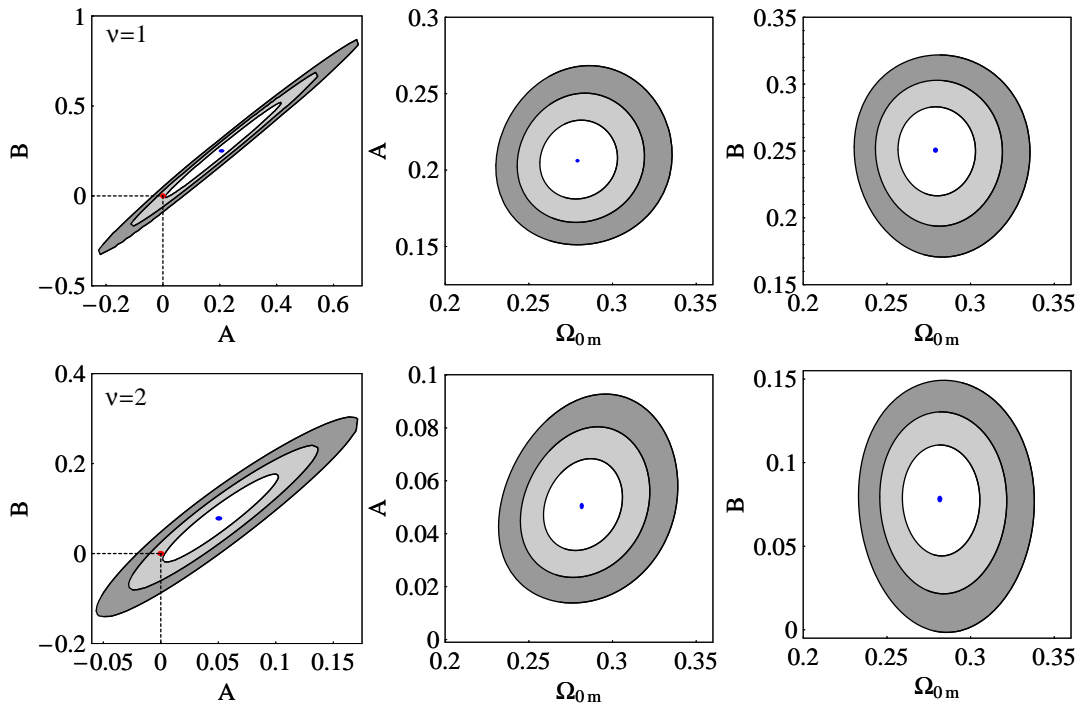}
\end{center}
\caption{\small $1\s, 2\s$ and $3\s$ contours in the parameter spaces $A - B$ (for best fit
$\O_{0m}$), $\O_{0m} - A$ (for best fit $B$), and $\O_{0m} - B$ (for best fit $A$), are shown
for the choices $\n =1$ (upper panels) and $\n = 2$ (lower panels). The best fit points for
both the choices are shown by the dots at the middle of all the $1\s$ contours, whereas the
cosmological constant, which corresponds to $A = B = 0$, is shown by the dot that is found
to lie on edge of the $1\s$ $A - B$ contour for both the choices (left panels, upper and
lower).}
\la{fig:2}
\end{figure}
%

\begin{figure}[!htb]
\begin{center}
\includegraphics[width=12.5cm,height=12.5cm]{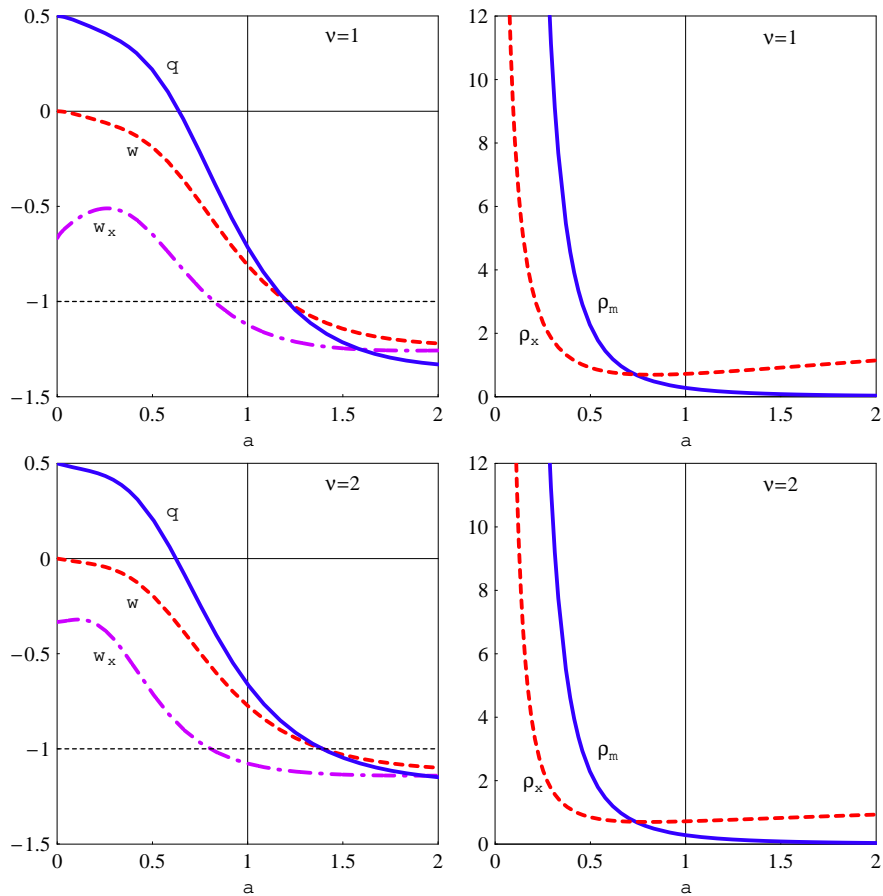}
\end{center}
\caption{\small Left panels: extrapolations of $\wx, w$ and $q$ (best fit with SN+CMB+BAO data),
as functions of the scale factor $a$, to the range $0 \leq a \leq 2$, for the choices $\n = 1$
(upper left) and $\n = 2$ (lower left). Right panels: extrapolated variations of $\rx$ and $\rmat$
(best fit with SN+CMB+BAO data) with the scale factor $a$, to the range $0 \leq a \leq 2$, for
$\n = 1$ (upper right) and $\n = 2$ (lower right). The $a = 1$ line denotes the present epoch.
The transition from a decelerating regime to an accelerating regime (i.e., the change of sign of
$q$) takes place at $a \sim 0.64$. The dark energy density $\rx$ is found to nearly follow the
matter density $\rmat$ for a considerable period in the past until becoming dominant very recently,
and increases slowly in the future.}
\la{fig:3}
\end{figure}

The $1\s, 2\s$ and $3\s$ contour plots of (i) $A$ versus $B$ (with $\O_{0m}$ fixed at its best fit
value), (ii) $\O_{0m}$ versus $A$ (with best fit $B$), and (iii) $\O_{0m}$ versus $B$ (with best
fit $A$), are shown in Fig. \ref{fig:2}, for the choices $\n =1$ (upper panels) and $\n = 2$
(lower panels). The case $A = B = 0$, which resembles a cosmological constant DE, is found to be
about $1\s$ away from the best fit point in the $A$ versus $B$ contours (left panels), for both
the choices.

The upper and lower left panels of Fig. \ref{fig:3} depict the variations of the best fit DE
EoS parameter $\wx (a)$, as well as the total EoS parameter $w (a)$, Eq. (\ref{tot-eos}), and
the deceleration parameter $q (a)$, Eq. (\ref{decel}), obtained as functions of the scale factor
$a$ (using the best fit values of the parameters $\O_{0m}, A, B$) and extrapolated to the range
$0 \leq a \leq 2$, for the choices $\n = 1$ and $\n = 2$ respectively. The range covers all of
the past, i.e., right from the big bang ($a = 0, z = \infty$) to the present ($a = 1, z = 0$),
and a considerable part in the future, up to $a = 2$ ($z = - 1/2$), i.e., when the present size
of the universe gets doubled. Both $\wx$ and $w$ are negative in the past and tend to become
constant at a value close to each other and a little less than $-1$ in the future. The value of
$q$, on the other hand, changes from positive to negative, i.e., the transition from deceleration
to acceleration takes place at $a = 0.640$ ($z = 0.562$) for $\n = 1$ and at $a = 0.624$ ($z =
0.603$) for $\n = 2$. In the future, $q$ also remains negative and tends to be steady at a value
close to $w$ and $\wx$. Thus the accelerated regime $q < 0$, as well as the `super-acceleration'
($\wx < -1$), do not appear to be transient in the present model.

The variations of the extrapolated best fit DE density $\rx$ and the matter density $\rmat$,
with the scale factor $a$ in the range $0 \leq a \leq 2$, are shown respectively for the choices
$\n = 1$ and $\n = 2$, in upper and lower right panels of Fig. \ref{fig:3}. For a considerable
period in the past the DE density nearly follows the the track of the matter density, until
exceeding the latter at scale factor $a \simeq 0.75$, and dominant thereafter. In other words,
$\rx$ decreases with $a$ in a similar manner as $\rmat$ does in the early regimes, until at a
recent epoch $a \simeq 0.75$, when the DE begins to dominate. This behaviour, although not
distinctly similar to that due to the tracker quintessence fields \cite{tracker}, may perhaps
stand as a possible resolution to the coincidence problem \cite{ss-prep2}. One can, in fact,
trace the similarity of the early universe profiles of $\rx$ and $\rmat$ to the form of the
chosen ansatz (\ref{f-ansz}) for the field solutions and the resulting expression (\ref{de-den3})
for $\rx$. In the early epochs, i.e., for small values of $a$, the DE density $\rx$ in Eq.
(\ref{de-den3}) is dominated by the inverse power-law term $\sim A a^{-\n}$, similar to the
matter density $\rmat = \O_{0m} a^{-3}$. However, since $\n < 3$ and the best fit value of $A$
is of the order of the best fit $\O_{0m}$, $\rx$ is smaller than $\rmat$, and decreases less
rapidly than the latter, for sufficiently smaller values of $a$. As $a$ increases, the value of
$\rx$ eventually exceeds $\rmat$ due to the presence of the positive constant term ($= \L$,
given by Eq. (\ref{CC})) in the expression (\ref{de-den3}) for $\rx$. The $B \ln a$ term in Eq.
(\ref{de-den3}), which is negative for $a < 1$ (i.e., past), is on the other hand, rather
sub-dominant compared to $\L$ and does not play a very significant role either in the past or
in near future. This is the reason why, the DE density $\rx$ increases slowly and does not
shoot up to very high values even at a scale factor as large as $a = 2$, giving rise to
singularities in finite future. Admittedly, of course $\rx \rightarrow \infty$ as $a \rightarrow
\infty$ due to the presence of the logarithmic term in $\rx$. Thus, the extrapolations of the
cosmological quantities using the best fit values of the model parameters, obtained in the
red-shift range $0 \leq z \leq 1.75$, appear to hold for very distant past and future.

In what follows, we integrate the expressions (\ref{sf-deriv1}) numerically in the next section
and use the values of $\O_{0m}, A$ and $B$ best fit with the data, so as to determine the variations
of the scalar fields $\f$ and $\xi$ with the scale factor $a$. We also reconstruct the potential $V$,
given in Eq. (\ref{pot2}), as a function of $a$, using these values of the parameters $\le(\O_{0m},
A, B\ri)$, and finally, we work out the approximate analytic expressions for the functional
variation of $V$ with $\f$ and $\xi$, in the regimes $a \ll 1$ (distant past) and $a \lesssim 1$
(recent past).

\section{Reconstruction of the scalar potential \la{sec:potential}}

Let us recall Eqs. (\ref{sf-deriv1}), from which one can derive the following equations for the
derivatives of the scalar fields $\f$ and $\xi$ with respect to the scale factor $a$:
\be \la{sf-deriv2}
\fl \qquad H_0 ~\f' (a) ~=~ \fr{\sq{A a^{-\n}} + \sq{A a^{- \n} + k}}{\sq{2}~ a ~\tH (a)} \, ,
\qquad H_0 ~\xi' (a) ~=~ \fr{\sq{A a^{-\n}} - \sq{A a^{- \n} + k}}{\sq{2}~ a ~\tH (a)} \, ,
\ee
where $\tH (a)$ is as given by Eq. (\ref{norm-hub2}) or (\ref{norm-hub3}), in terms of the model
parameters $\le(\O_{0m}, A, B\ri)$.

Assuming the initial condition that $\f = \xi = 0$ at $a = 0$, one may re-write the above equations
in integral form as
\be \la{sf-sol}
H_0 ~\f (a) ~=~ \sq{\fr A 2} \le[I_+ (a) - I_+ (0)\ri] \, ,
\qquad H_0 ~\xi (a) ~=~ \sq{\fr A 2} \le[I_- (a) - I_- (0)\ri] \, ,
\ee
where
\be \la{Ipm}
I_{\pm} (a) ~=~ \int^a \fr{d\ta}{\ta^{(1 + \n/2)} \tH (\ta)}  \le[1 ~\pm~ \sq{1 ~+~ \fr{2 k \ta^\n} A}\ri] .
\ee
Again, denoting $\f = \f_0$ and $\xi = \xi_0$ at the present epoch ($a = 1$), we have
\be \la{sf-present}
H_0 ~\f_0 ~=~ \sq{\fr A 2} \le[I_+ (1) - I_+ (0)\ri] \, ,
\qquad H_0 ~\xi_0 ~=~ \sq{\fr A 2} \le[I_- (1) - I_- (0)\ri] \, .
\ee
From Eqs. (\ref{sf-sol}) and (\ref{sf-present}), one therefore finds
\be \la{sf}
\fr{\f (a)}{\f_0} ~=~ \fr{I_+ (a) - I_+ (0)}{I_+ (1) - I_+ (0)} \, ,
\qquad \fr{\xi (a)}{\xi_0} ~=~ \fr{I_- (a) - I_- (0)}{I_- (1) - I_- (0)} \, .
\ee
%

\begin{figure}[!htb]
\begin{center}
\includegraphics[width=12.5cm,height=12.5cm]{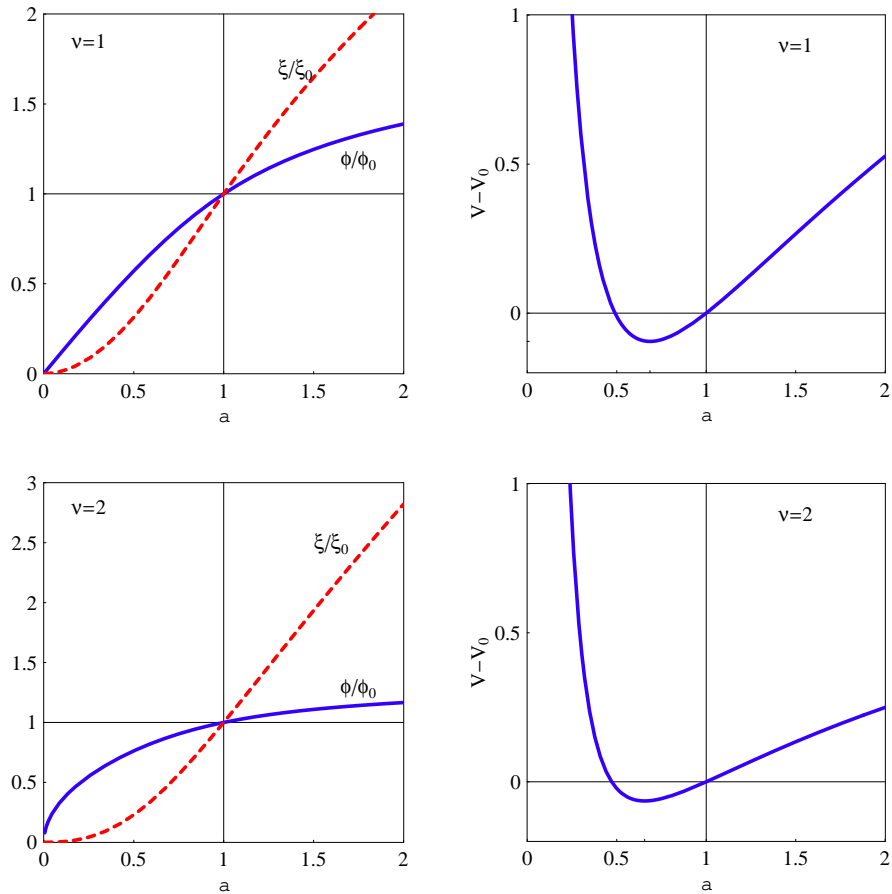}
\end{center}
\caption{\small Evolutions of the normalized scalar fields $\f/\f_0$ and $\xi/\xi_0$ (left panels),
and the scalar potential $V$ minus its present value $V_0$ (right panels), with the scale factor $a$,
for the choices $\n = 1$ and $\n = 2$. Such evolutions, which are reconstructed using the best fit
values of the model parameters $\O_{0m}, A$ and $B$, have been extrapolated to the range $0 \leq a
\leq 2$. The vertical line $a = 1$ resembles the present epoch.}
\la{fig:4}
\end{figure}

Now, in order to perform the integrations in $I_\pm$, Eq. (\ref{Ipm}), one has to assign a particular
value to the parameter $k$. Eliminating $k$ from Eqs. (\ref{Q-sq}) and (\ref{B-def}) we find that the
kinetic interaction $Q$ satisfies the following cubic equation involving the parameters $\b, \c$ and
$B$:
\be \la{Q-eqn}
4 Q^3 ~-~ \b \c ~ Q^2 ~+~ 2 \le(\b B - 2\ri) Q ~+~ \b \c ~=~ 0 \, .
\ee
Solving this equation and substituting the feasible root\footnote{Feasibility here implies that one
should pick only that root $Q$ which is real and greater than unity, so that the presumption of the
positivity of $k$ is ensured through the relation (\ref{Q-sq}).} back in the relation (\ref{Q-sq}),
one finds $k$ in terms of $\b, \c$ and $B$ only. For $B$, we use its values best fit with the
observational data, given in table 1 for the choices $\n = 1$ and $\n = 2$. For $\b$ and $\c$, we
recall that their values should be such that the condition $\c^{-1} \ll \b \ll 1$ is satisfied and
the KIDQ Lagrangian (\ref{de-action1}) could emerge as an approximation to the two-field DBI
Lagrangian (see the Appendix). Henceforth, assuming typically $\b = 0.01$ and $\c = 10^4$, we find
$k = 0.012$ for $\n = 1$ and $k = 0.004$ for $\n = 2$. Also, using the best fit values of the
parameters $\O_{0m}, A$ and $B$, on which $\tH$ depends, and performing numerically the integrations
in $I_\pm$, Eq. (\ref{Ipm}), we finally determine the variations of the normalized scalar fields
$\f/\f_0$ and $\xi/\xi_0$ with the scale factor $a$, for $\n = 1$ and $\n = 2$. Similarly, we also
find how the quantity $V - V_0$, given by Eq. (\ref{pot2}), varies with $a$, for the same choices
of $\n$. Such variations, extrapolated to the range $0 \leq a \leq 2$, are shown in Fig. \ref{fig:4}.
We observe that both $\f/\f_0$ and $\xi/\xi_0$ increases with increasing $a$, however $\xi/\xi_0$
grows much faster than $\f/\f_0$, whose variation gradually decreases with $a$ (see the left panels
of Fig \ref{fig:4}). As such the ratio $\xi/\xi_0$, which has been less than $\f/\f_0$ in the past
(i.e., $a < 1$), becomes greater than $\f/\f_0$ for $a > 1$ and grows to high values as we
extrapolate it to far future. The potential $V$, on the other hand, decreases from a very high
value ($\gg V_0$, its present value) in the early epochs, reaches a minimum ($< V_0$) at some point
$a = a_m$ in the past, and increases steadily thereafter. However, the entire profile of $V - V_0$
for both $\n = 1$ and $\n = 2$ (shown in the right panels of Fig \ref{fig:4}), is not symmetric
about the minimum value $V_m - V_0$. In fact, the asymmetry is more when $\n = 2$, rather than when
$\n = 1$. That is, the potential, after reaching its minimum, increases rather slowly for greater
values of $\n$. The values of $a_m$ and $V_m$ (not shown in Fig. \ref{fig:4}) could be calculated
by extremizing the expression (\ref{pot2}) for $V$:
\bea \la{min}
\fl \qquad a_m ~= \le[\fr{\le(6 - \n\ri) A}{6 ~B}\ri]^{1/\n} \, , \qquad
V_m ~=~ V_0 ~-~ \fr{\le(6 - \n\ri) A}{\n} ~+~ \fr{6 ~B}{\n} ~\ln \le[\fr{\le(6 - \n\ri) A}{6 ~B}\ri] .
\eea
For $\n = 1$, $a_m = 0.6856$ and $V_m - V_0 = - 0.0948$, whereas for $\n = 2$, $a_m = 0.6558$ and
$V_m - V_0 = - 0.0643$. Therefore, the greater the value of $\n$, the earlier is the occurance of the
minimum in the past, and the lesser is its value in magnitude.

The overall variation of the potential $V$ with the scale factor $a$ could be explained as follows:
In the very early epochs $V - V_0$, given by Eq. (\ref{pot2}), is very large and positive due to the
dominance of the positive inverse power-law term $(6/\n - 1) A a^{-\n}$. With the increase of $a$,
this term rapidly diminishes, and $V - V_0$ becomes negative when the term $6 B \ln a$, which is
negative for $a < 1$ (past), starts to dominate over the positive second term on the right hand side
of Eq. (\ref{pot2}). Eventually, $V - V_0$ reaches the minimum, and then the potential $V$ starts
to increase as the term $6 B \ln a$, though negative, gradually decreases in magnitude. $V$ becomes
equal to $V_0$ at the present epoch ($a = 1$) and after that $V - V_0$ increases with positive
values as the logarithmic term becomes positive and increases with $a$. The asymmetry of the two
sides of the minimum is obvious, because one is due to a power-law fall off and the other is due
to a logarithmic increment. Also, for a bigger value of $\n$ (here $\n = 2$), the power-law fall
off is faster. The asymmetry is therefore more distinct, the minimum is attained earlier, and the
minimum value $V_m$ is smaller in magnitude.

To reconstruct the potential $V$ as a function of the fields $\f$ and $\xi$, we need to solve the
Eqs. (\ref{sf-deriv2}) (or, equivalently need to work out the integrals $I_\pm$, Eq. (\ref{Ipm}))
analytically. However, this is very difficult because of the fairly complicated form of the
normalized Hubble parameter $\tH$, given by Eq. (\ref{norm-hub3}). As an alternative, we resort
to the following two regimes which are relevant for us: (i) $a \ll 1$ (early past) and (ii) $a
\simeq 1$ (recent past, present, and near future), and work out the approximate functional form
of $V (\f, \xi)$ in these regimes.

\bigskip
\noindent
{\bf (i) Early Universe:}

\bigskip

For $a \ll 1$, the Hubble expansion is dominated by the inverse power-law terms in Eq.
(\ref{norm-hub2}). As such, one can approximate:
\be \la{norm-hub-app1}
\tH^2 (a) ~\approx~ \fr{6 A}{\n a^\n} ~+~ \fr{\O_{0m}}{a^3} \, .
\ee
Now, from Eqs. (\ref{sf-sol}) and (\ref{Ipm}), we have
\be \la{sf-sum1}
H_0 \le[\f (a) ~+~ \xi (a)\ri] =~ \sq{2 A} \le[I (a) ~-~ I (0)\ri] ,
\ee
where
\be \la{I}
I (a) ~=~ \fr 1 2 \le[I_+ (a) + I_- (a)\ri] =~ \int^a \fr{d\ta}{\ta^{1 + \n/2} \tH (\ta)} \, .
\ee
Using the approximated form (\ref{norm-hub-app1}) of $\tH$, we get
\be \la{I-app1}
I (a) ~\approx~ \fr 1 {3 - \n} \sq{\fr{2 \n}{3 A}}~ \sinh^{-1} \le(\sq{\fr{6 A}{\n \O_{0m}}} ~
a^{(3 - \n)/2}\ri)  \qquad \Rightarrow \quad I (0) \approx 0 \, ,
\ee
whence
\be \la{sf-sum1-app1}
H_0 \le[\f (a) ~+~ \xi (a)\ri] \approx~ \fr 2 {3 - \n} \sq{\fr{\n} 3}~ \sinh^{-1}
\le(\sq{\fr{6 A}{\n \O_{0m}}} ~ a^{(3 - \n)/2}\ri) .
\ee
Inverting this relation and substituting in Eq. (\ref{pot2}), one finally obtains
\bea \la{V-app1}
\fl \quad V (\f, \xi) ~\approx~ V_0 &+& \le(\fr 6 {\n} - 1\ri) A \le(\fr{\n \O_{0m}}
{6 A}\ri)^{-\n/(3 - \n)} \sinh^{-2\n/(3 - \n)} \le[\fr{3 - \n} 2 \sq{\fr 3 {\n}} ~
H_0 \cdot \le(\f + \xi\ri)\ri] \nn \\
&+& \fr{12 B}{3 - \n} ~\ln \le\{\sinh \le[\fr{3 - \n} 2 \sq{\fr 3 {\n}} ~ H_0 \cdot
\le(\f + \xi\ri)\ri]\ri\} \nn \\
&-& \le(\fr 6 {\n} - 1\ri) A ~+~ \fr{6 B}{3 - \n} ~\ln \le(\fr{\n \O_{0m}}{6 A}\ri) .
\eea

\begin{figure}[!htb]
\begin{center}
\includegraphics[width=12.5cm,height=12.5cm]{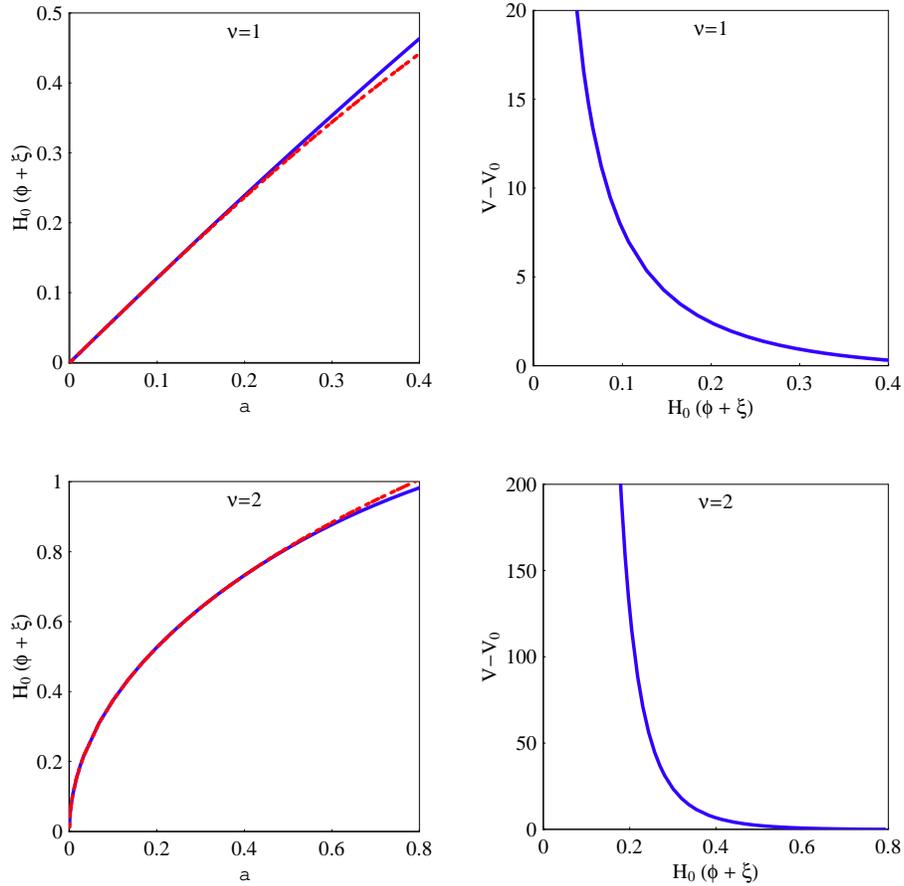}
\end{center}
\caption{\small Exact (solid line) and approximated (dashed line) variations of the
quantity $H_0 (\f + \xi)$ with the scale factor $a$ in the early epochs ($a \ll 1$),
for the choices $\n = 1$ (upper left) and $\n = 2$ (lower left). The approximation
is found to hold up to  $a \sim 0.3$ for $\n = 1$, and up to $a \sim 0.7$ for $\n =
2$. The right panels show the variation of the approximated $V - V_0$ with $H_0 (\f
+ \xi)$ for the range of validity of the approximation, in the cases $\n = 1$ and
$\n = 2$ (upper right and lower right, respectively).}
\la{fig:5}
\end{figure}

The left panels of Fig. \ref{fig:5} show how the approximated form of the quantity
$H_0 (\f + \xi)$, as well as its exact form (obtained by working out the integral
$I$, Eq. (\ref{I}), numerically), vary with the scale factor $a$, for the choices
$\n = 1$ (upper left) and $\n = 2$ (lower left). Whereas for $\n = 1$, the
approximation is found to be valid only up to $a \sim 0.3$, it holds good till
$a \sim 0.7$ for $\n = 2$. Within the region of validity of the approximation,
$H_0 (\f + \xi)$ increases almost linearly with $a$ for $\n = 1$, whereas for
$\n = 2$, $H_0 (\f + \xi)$ increases but gradually slows down as $a$ increases.
The variation of the approximated $V - V_0$ as a function of the fields $\f$ and
$\xi$, Eq. (\ref{V-app1}), is shown for $\n = 1$ and $\n = 2$ in the upper right
and lower right panels of Fig. \ref{fig:5} respectively. Both these plots extend
up to the range of $a$ for which the approximation is valid in respective cases.
The potential varies smoothly (i.e, without any discontinuity or multi valued-ness)
with $H_0 (\f + \xi)$, as with the scale factor $a$ (in Fig. \ref{fig:4}). Also
since  $H_0 (\f + \xi)$ increases monotonically with $a$, the nature of the $V -
V_0$ versus $H_0 (\f + \xi)$ plots in Fig. \ref{fig:5} is similar to the nature of
the $V - V_0$ versus $a$ plots in Fig. \ref{fig:4} for smaller values of $a$.

\bigskip
\noindent
{\bf (ii) Recent Universe:}

\bigskip

Expanding the expression (\ref{norm-hub3}) for $\tH$ in powers of $(1 - a)$, for
$a \approx 1$ (i.e., close to the present epoch), and retaining only the terms
linear in $(1 - a)$ we have
\be \la{norm-hub-app2}
\tH^2 (a) ~\approx~ 1 ~+~ \tilh \le(1 - a\ri) \, , \qquad \mbox{where} \quad
\tilh = 6 \le(A - B + \fr{\O_{0m}} 2\ri) ,
\ee
for both $\n = 1$ and $\n = 2$. Now, from Eqs. (\ref{sf-sol}), (\ref{Ipm}) and
(\ref{sf-present}), one obtains\footnote{Note that Eq. (\ref{sf-sum1}) cannot be
used now, because the approximation does not hold for $a = 0$.}
\be \la{sf-sum2}
H_0 \le[\le(\f (a) - \f_0\ri) + \le(\xi (a) - \xi_0\ri)\ri] =~ \sq{2 A} \le[I (a)
~-~ I (1)\ri] ,
\ee
where $I (a)$ is the integral given by Eq. (\ref{I}). One can evaluate $I (a)$
numerically for the choices $\n = 1$ and $\n = 2$ and find $H_0 \le[(\f - \f_0)
+ (\xi - \xi_0)\ri]$, using the best values of the parameters $\O_{0m}, A$ and
$B$. However, to determine the functional form $V (\f, \xi)$, we need to work
out the integral $I (a)$ analytically. Let us separately consider the cases
$\n = 1$ and $\n = 2$ as follows:

\bigskip
\noindent
\underline{For $\n = 1$:}
The expression (\ref{pot2}) for the potential $V$ can be approximated as
\be \la{pot2-app1}
V (a) ~=~ V_0 ~+ \le(5 A - 6 B\ri) \le(1 - a\ri) .
\ee
On the other hand, the approximate analytic evaluation of the integral $I (a)$,
Eq. (\ref{I}), leads to
\be \la{sf-sum2-app1}
H_0 \le[\le(\f (a) - \f_0\ri) + \le(\xi (a) - \xi_0\ri)\ri] \approx~ - \fr{2
\sq{2 A}}{1 + \tilh} \le[\le(\fr{1 + \tilh} a - \tilh\ri)^{1/2} ~-~ 1\ri] .
\ee
Inverting this expression and substituting in the above equation (\ref{pot2-app1}),
we get
\bea \la{V-app21}
\fl \qquad V (\f, \xi) ~\approx~ V_0 ~+ \le(5 A - 6 B\ri) \le[1 ~-~ \fr{1 +
\tilh}{\le\{1 - \fr{\le(1 + \tilh\ri) H_0}{2 \sq{2 A}} \le[(\f - \f_0) +
(\xi - \xi_0)\ri]\ri\}^2 + \tilh}\ri] .
\eea
%

\begin{figure}[!htb]
\begin{center}
\includegraphics[width=12.5cm,height=12.5cm]{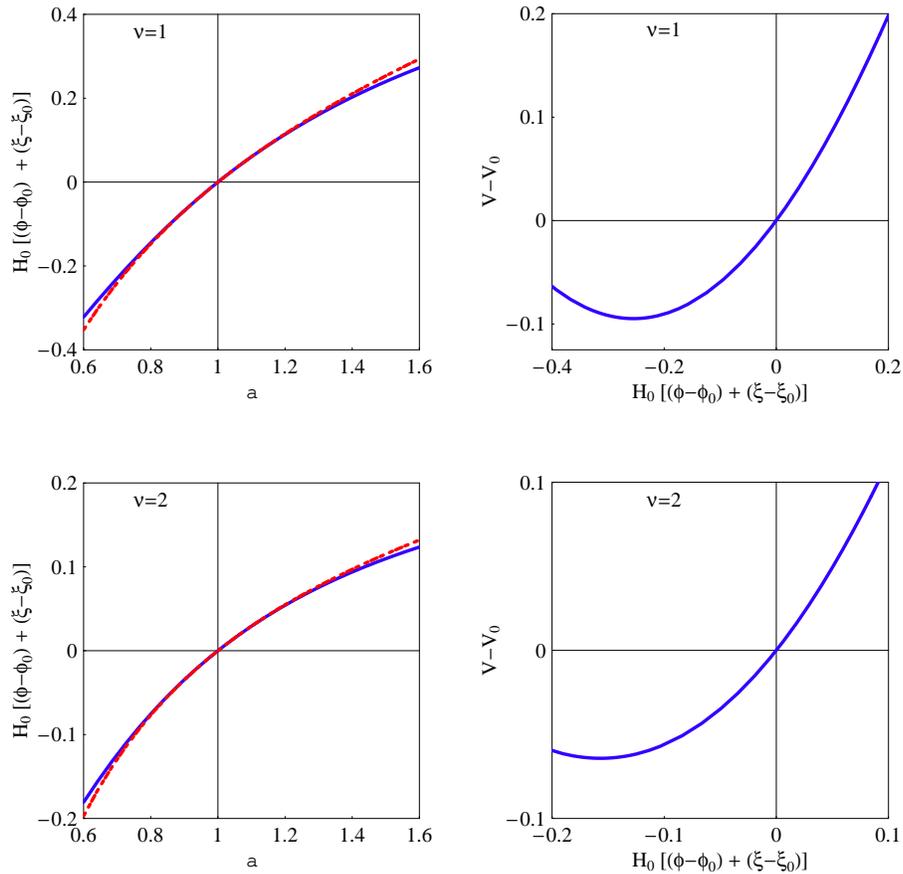}
\end{center}
\caption{\small Exact and approximate (solid and dashed lines) variations of $H_0
\le[(\f - \f_0) + (\xi - \xi_0)\ri]$ with the scale factor $a (\approx 1$), as well
as the approximated functional variation $\le[V (\f, \xi) - V_0\ri]$, are shown for
the choices $\n = 1$ (upper panels) and $\n = 2$ (lower panels). The approximation
is found to be good in the range $0.7 \lesssim a \lesssim 1.4$ for both $\n = 1$ and
$2$. Within this range $V - V_0$ has a minimum at $H_0 \le[(\f - \f_0) + (\xi -
\xi_0)\ri] \simeq -0.25$ (for $\n = 1$) and $\simeq -0.15$ (for $\n = 2$).}
\la{fig:6}
\end{figure}

The variations of the exact and approximated forms of $H_0 \le[(\f - \f_0) + (\xi -
\xi_0)\ri]$ with the scale factor $a$, as well as the functional variation of the
approximated $\le[V (\f, \xi) - V_0\ri]$, are shown in the upper panels (left and
right respectively) of Fig. \ref{fig:6}. The approximation is found to hold for a
fairly large range $0.7 \lesssim a \lesssim 1.4$. The approximated $\le[V (\f, \xi)
- V_0\ri]$, which has been plotted for this range of validity, has a minimum ($\simeq
-0.1$) at $H_0 \le[(\f - \f_0) + (\xi - \xi_0)\ri] \sim -0.25$. The overall profile
of the approximated $V - V_0$ is similar to the exact variation of $V - V_0$ with $a$
(shown in Fig. \ref{fig:4}) in the range $0.7 < a < 1.4$ where the approximation is
found to be valid.

\bigskip
\noindent
\underline{For $\n = 2$:}
The approximation of the expression (\ref{pot2}) for the potential $V$ is given by
\be \la{pot2-app2}
V (a) ~=~ V_0 ~+~ 2 \le(2 A - 3 B\ri) \le(1 - a\ri) ,
\ee
and by evaluating the integral $I (a)$ given by Eq. (\ref{I}), approximately, one
finds
\bea \la{sf-sum2-app2}
\fl H_0 \le[\le(\f (a) - \f_0\ri) + \le(\xi (a) - \xi_0\ri)\ri] \approx~ \sq{\fr{2 A}
{1 + \tilh}} \le[\le\{\sq{1 - \fr{\tilh}{1 + \tilh}} ~-~ \fr 1 a \sq{1 - \fr{\tilh a}
{1 + \tilh}}\ri\} \ri. \nn \\
+ \le. \fr{\tilh}{1 + \tilh} \le\{\tanh^{-1} \le(\sq{1 - \fr{\tilh}{1 + \tilh}}\ri)
-~ \tanh^{-1} \le(\sq{1 - \fr{\tilh a} {1 + \tilh}}\ri)\ri\} \ri] .
\eea
This expression cannot be inverted to get the scale factor $a$ as a function of $H_0
\le[(\f - \f_0) + (\xi - \xi_0)\ri]$, and hence obtain the expression for the analytic
functional variation of the approximated $V (\f, \xi)$. However, one may find the
parametric plot of the approximated $\le[V (\f, \xi) - V_0\ri]$ versus $H_0 \le[(\f
- \f_0) + (\xi - \xi_0)\ri]$ using the above equations (\ref{pot2-app2}) and
(\ref{sf-sum2-app2}). Such a plot is shown in the lower right panel of Fig. \ref{fig:6}.
$\le[V (\f, \xi) - V_0\ri]$ varies in a similar way as for $\n = 1$ (see the upper
right panel of Fig. \ref{fig:6}), however the minimum value is higher and the minimum
is reached at a value $H_0 \le[(\f - \f_0) + (\xi - \xi_0)\ri] \simeq -0.15$, greater
than that for $\n = 1$. The lower left panel of Fig. \ref{fig:6}) shows how the exact
and approximate forms of $H_0 \le[(\f - \f_0) + (\xi - \xi_0)\ri]$ vary with the scale
factor $a$. Similar to the case $\n = 1$ (upper left), we find that the approximation
remains valid in a fairly large region $0.7 \lesssim a \lesssim 1.4$ as well for $\n
= 2$ (lower left).

Finally, it should be mentioned here that although we have expressed the potential $V$
approximately as a function of $(\f + \xi)$ for both $a \ll 1$ and $a \simeq 1$ (see
the Eqs. (\ref{V-app1}) and (\ref{V-app21})), strictly speaking this cannot be true at
all epochs. Indeed, if $V$ is exactly a function of $(\f + \xi)$, i.e., $\pa V/\pa \f
= \pa V/\pa \xi$, then the ansatz (\ref{f-ansz}) becomes inconsistent with the scalar
field equations of motion (\ref{sf-eom}), given in sec. \ref{sec:model}, and all our
above results are invalid. The exact form of the potential $V$, which we are studying
in a later work (in preparation) \cite{ss-prep2}, should therefore be asymmetric in
$\f$ and $\xi$ (at least to their linear order) in order the model to be consistent.

\section{Conclusions  \la{sec:concl}}

We have thus explored the plausible crossing of the cosmological constant ($\L$) barrier
by the dark energy equation of state parameter $\wx$ in a fairly simple set-up of two
canonical (quintessence-type) scalar fields with a mutual kinetic interaction. Such a
crossing, which has been a big problem in many scalar field DE models, is shown to be
realized with a specific form of the kinetic interaction and with the requirement that
the dynamical part of total DE Hamiltonian is positive definite, so that the model is
quantum mechanically consistent. Classical stability of the model is also guaranteed as
the squares of the sound speeds corresponding to the adiabatic and entropy perturbation
modes are positive definite.

Under certain limiting conditions, the specific form of the kinetic interaction, which we
study, can be shown to have originated from a higher dimensional two-scalar DBI action,
that appears in the string theoretic scenario \cite{dbi}. Such a kinetic interaction
provides additional flexibility in $\wx$ (apart from those provided by the usual kinetic
terms of the scalar fields), so that the $\L$-barrier (i.e., $\wx = -1$) could be crossed
at a particular epoch.

Joint constraints on the parameters of the model by the SN+CMB+BAO data due to the SN
Search Team \cite{kowal}, WMAP \cite{wmap5} and SDSS \cite{sdss}, show that $\wx$ has
most likely crossed $-1$ at a recent red-shift $z = z_c$ ($0.215 \leq z_c \leq 0.245$),
and its value $w_{_{0X}}$ at present is less than $-1$ ($-1.123 \leq w_{_{0X}} \leq
-1.077$). On the other hand, the transition from the decelerated phase of expansion of
the universe to the accelerated phase takes place between $0.562 < z < 0.603$. All these
results are fairly consistent with the model-independent estimates with the SN+CMB+BAO
data in ref. \cite{wmap5}. Additionally, we also observe that the dark energy density
(best fit with the observational data) nearly follows the matter density (i.e., exhibits
a similar fall-off with the scale factor $a$, as the latter) at early epochs, until
exceeding it very recently. This apparently could provide a resolution to the coincidence
problem (that is associated with the cosmological constant). Extrapolations to future
epochs also show that the best fit dark energy density increases fairly slowly even at
a fairly large scale factor, implying that singularities in finite future may plausibly
be avoided in our model.

The numerical reconstruction of functional forms of the scalar fields and the scalar
potential, using the best fit values of the model parameters, shows smooth variations
with the scale factor $a$, although analytically the exact form of the potential as
function of the fields is very difficult to obtain. Working out therefore the approximate
solutions for the scalar fields in the early universe and near the present epoch,
we have obtained approximate analytic functional forms of the potential in terms of
the scalar fields, in these regimes. Such analytic forms also exhibit the same smooth
nature as the numerically reconstructed potential.

Some interesting questions that arise in the context of the present model are in order:
\begin{itemize}
\item Can one unify dark matter and dark energy in the general framework of a kinetically
interacting double or multi-scalar theory, instead of treating them separately as in this
model?

\item Can we generically determine for kinetically interacting double or multi-quintessence
model, the exact form of the scalar potential,  which could lead to the $\L$-barrier crossing
as well as the dark matter tracking (by DE)? If so, then could it be ascertained whether such
a potential belongs the class of tracking or scaling potentials that arise in generic k-essence
theories?

\item Can we have in the context of a kinetically interacting double or multi-quintessence
model, the assisted accelerated solutions, which have been shown to exist generically for
the multi-field k-essence models admitting scaling solutions \cite{tsuji}?

\item Can we ascertain the status of the future singularities, if any, in the context of
kinetically interacting double-quintessence models, generically, i.e., by not just resorting
to a particular ansatz to solve for the field equations? Or, can we generically ascertain
whether or not the cosmic super-acceleration ($\wx < -1$ regime) is always eternal (as in
the present model) for kinetically interacting two-field quintessence?
\end{itemize}
Works addressing some of these questions are in progress \cite{ss-prep2}, which we hope to
report soon.

\bigskip

\ack{The author acknowledges useful discussions with the members of the theoretical physics
group of University of Lethbridge, and especially to Saurya Das for many helpful remarks
and suggestions. This work is supported by the Natural Sciences and Engineering Research
Council of Canada.}


\section*{Appendix: Kinetic interaction from a two-field DBI perspective}

The Dirac-Born-Infeld (DBI) multi-scalar Lagrangian is particularly important in view of
the notion acquired from string theory that our observable four dimensional world may be
looked upon as being a warped D3-brane embedded in a higher dimensional (bulk) space-time
\cite{kirit}. The general expression for such a Lagrangian in an effective four dimensional
theory is given by \cite{lang1,lang2}:
\be \la{dbi-lag1}
P ~=~ \c \le(1 ~-~ \sq{\mathcal{D}}\ri)  -~ U (\f^I) \, ,
\ee
where $\f^I (I = 1,2,\dots)$ are a set of scalar fields, which from the bulk point of
view, correspond to the coordinates of the brane in the extra dimensions, and $U (\f^I)$
is the multi-field potential due to the interaction of the brane with bulk fields or with
fields on other branes. $\mathcal{D}$ is the determinant of induced metric on the brane,
given by
\be \la{D}
\mathcal{D} ~=~ \mbox{det} \le(\d_\m^\n ~+~ \c^{-1}~ \mathcal{G}_{IJ} ~\pa_\m \f^I
\pa^\n \f^J\ri) \, ,
\ee
where $\mathcal{G}_{IJ}$ is the field space metric, which is proportional to the extra
dimensional metric living in the bulk, and $\c$ is a coupling parameter, that appears by
virtue of the warping of the D3-brane in the bulk. Both $\mathcal{G}_{IJ}$ and $\c$ could
generally be functions of the fields $\f^I$, and in a homogeneous FRW background the above
expression for $\mathcal{D}$ takes the form \cite{lang1,lang2}:
\be \la{D0}
\mathcal{D} ~=~ 1 ~-~ 2 \c^{-1} ~\mathcal{G}_{IJ} ~ X^{IJ} \, , \qquad
X^{IJ} ~=~ \fr{\df^I \df^J} 2 \, .
\ee

Now, for a configuration of two fields $\f^I := \le\{\f, \xi\ri\}$, assuming $\c$ and
$\mathcal{G}_{IJ}$ to be constants (for all $I,J$), we can write
\be \la{D1}
\mathcal{D} ~=~ 1 ~-~ \c^{-1} \le(\mathcal{G}_{11}~ \df^2 ~+~ \mathcal{G}_{22}~ \dot{\xi}^2
~+~ 2 \mathcal{G}_{12}~ \df \dot{\xi}\ri) \, .
\ee
Rescaling $\f$ and $\xi$ such that both the metric components $\mathcal{G}_{11}$ and
$\mathcal{G}_{22}$ are effectively set to unity, we get
\be \la{D2}
\mathcal{D} ~=~ 1 ~-~ \fr{\df^2 + \dot{\xi}^2}{\c} ~-~ \fr{\b} 2 \df \dot{\xi} \, ,
\qquad \mbox{where} \quad \b = \fr{4 \mathcal{G}_{12}}{\c} = \mbox{constant} \, .
\ee

Under the assumption:
\be \la{parameters}
\c \gg \mathcal{G}_{12} \gg 1 \, \qquad \Rightarrow \quad \b \ll 1 \, , \qquad \mbox{but} \quad
\c^{-1} \ll \b \, ,
\ee
one can write\footnote{Note that in this paper we have typically set $\c = 10^4$ and with
$\mathcal{G}_{12} = 10^2$, it implies $\b = 4 \mathcal{G}_{12}/\c = 10^{-2} \ll 1$ (but
$\gg \c^{-1}$), as per the assumption (\ref{parameters}) [see sec. \ref{sec:potential}].}
\be \la{D-sq}
\sq{\mathcal{D}} ~=~ -~ \fr{\df^2 + \dot{\xi}^2}{2 \c} ~+~ \sq{1 ~-~ \fr{\b} 2 \df \dot{\xi}}
~+~ \mathcal{O} \le(\fr{\b}{\c}\ri) \, .
\ee
Substituting this in Eq. (\ref{dbi-lag1}), and neglecting the $\mathcal{O} \le(\b/\c\ri)$ and
higher order terms, we finally obtain
\be \la{KIDQ-lag}
P ~=~ \fr{\df^2 + \dot{\xi}^2} 2 ~-~ \c ~\sqrt{1 ~-~ \fr{\b} 2 \df \dot{\xi}} ~-~
V (\f, \xi) \, .
\ee
This is the same as the KIDQ Lagrangian (\ref{de-action1}) considered in sec. \ref{sec:model};
$V (\f, \xi) = U (\f, \xi) - \c$ being the effective (shifted) two-scalar potential.

\bigskip
\noindent
{\bf Stability criterion:}
In general multiple scalar field models, the cosmological perturbations of scalar type are
divided into: (i) the adiabatic (instantaneous) modes, which are fluctuations along the
field space trajectory, and (ii) the entropy modes, which are orthogonal to the field space
trajectory \cite{gordon}. The squares of the speeds of propagation of both these modes
should be positive definite in order that the underlying model is cosmologically stable.
%

\begin{figure}[!htb]
\begin{center}
\includegraphics[width=12.5cm,height=6.25cm]{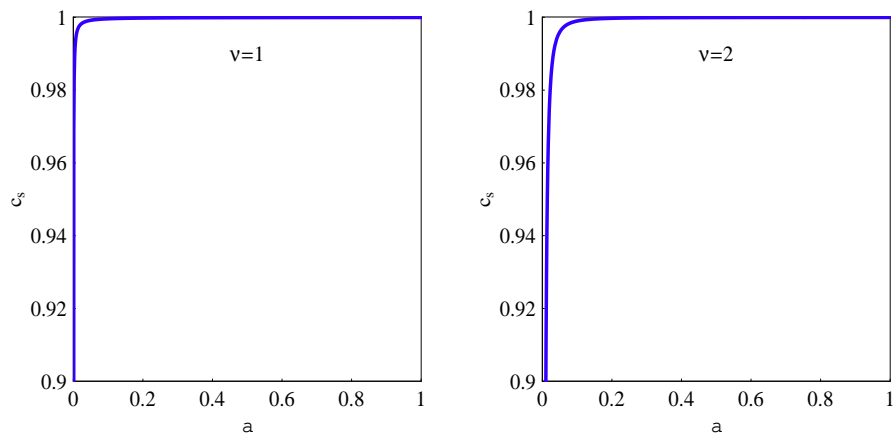}
\end{center}
\caption{\small Plots of the KIDQ sound speed $c_s$ (assumed to be isotropic) as a function
of the scale factor $a$, for the choices $\n = 1$ and $\n = 2$. Except at very early epochs
($a \ll 1$), the sound speed is found be extremely close to unity (i.e., the speed of light),
for both the choices.}
\la{fig:7}
\end{figure}

For multi-scalar DBI models \cite{lang1,lang2}, the adiabatic and entropy modes are shown
to be isotropic, i.e., they propagate with the same (sound) speed $c_s$ equal to
$\sq{\mathcal{D}}$. Since the KIDQ Lagrangian is an approximation of the two-scalar DBI
Lagrangian under the assumption (\ref{parameters}), one may intuitively consider this
isotropy of the perturbation modes exists in the case of KIDQ as well, and that the
effective sound speed for KIDQ to be $c_s = \sq{\mathcal{D}}$, given by Eq. (\ref{D-sq}).
Obviously, as shown in Fig. \ref{fig:7}, this sound speed (and as such its square) is
positive definite and very close to unity (speed of light) for both $\n = 1$ and $\n = 2$,
since the $\c^{-1}$ and $\b$ terms in Eq. (\ref{D-sq}) are very small compared to unity.

More generally of course, if we treat KIDQ as an exact model (i.e., not an approximation
of DBI), then a rigorous analysis following the general formalism worked out in \cite{lang2},
shows that the sound speeds corresponding to the adiabatic and entropy modes do actually
differ. However, the difference is very slight for the values of the parameters $\b, \c$
used in this paper, and the sound speeds are still positive definite and close to unity
\cite{ss-prep2}.

\end{document}